\makeatletter                   %1
\def\mdseries@tt{m}             %1
\makeatother                    %1

\documentclass[sigplan]{acmart}

\usepackage{color}
\hypersetup{
    colorlinks=true,
    linkcolor=blue,
    filecolor=red,      
    urlcolor=magenta,
    breaklinks=true,            %3
}
\usepackage{breakurl}           %3

\acmConference[TyDe'19]{ACM SIGPLAN Workshop on Type-driven Development}{August 18, 2019}{Berlin, Germany}
\acmYear{2019}
\acmISBN{} % \acmISBN{978-x-xxxx-xxxx-x/YY/MM}
\acmDOI{} % \acmDOI{10.1145/nnnnnnn.nnnnnnn}
\startPage{1}

\setcopyright{none}
\usepackage{booktabs}   %% For formal tables:
\usepackage{subcaption} %% For complex figures with subfigures/subcaptions
\usepackage{algorithm2e}

\usepackage[frozencache,cachedir=.]{minted}

\usepackage{lipsum}

\usepackage{float}

\begin{document}
\sloppy

%% Title information
\title[Augmenting Type Signatures for Program Synthesis]{Augmenting Type
Signatures for Program Synthesis}
% \titlenote{(Extended Abstract)}             %% \titlenote is optional;
                                        %% can be repeated if necessary;
                                        %% contents suppressed with 'anonymous'
\subtitle{(Extended Abstract)}                     %% \subtitle is optional
% \subtitlenote{with subtitle note}       %% \subtitlenote is optional;
                                        %% can be repeated if necessary;
                                        %% contents suppressed with 'anonymous'

%% Author information
%% Contents and number of authors suppressed with 'anonymous'.
%% Each author should be introduced by \author, followed by
%% \authornote (optional), \orcid (optional), \affiliation, and
%% \email.
%% An author may have multiple affiliations and/or emails; repeat the
%% appropriate command.
%% Many elements are not rendered, but should be provided for metadata
%% extraction tools.

%% Author with single affiliation.
\author{Bruce Collie}
\affiliation{
  % \department{School of Informatics}
  \institution{University of Edinburgh}
  % \streetaddress{10 Crichton Street}
  % \city{Edinburgh}
  % \state{Scotland}
  % \postcode{EH8 9AB}
  % \country{UK}
}
\email{bruce.collie@ed.ac.uk}

\author{Michael O'Boyle}
\affiliation{
  % \department{School of Informatics}
  \institution{University of Edinburgh}
  % \streetaddress{10 Crichton Street}
  % \city{Edinburgh}
  % \state{Scotland}
  % \postcode{EH8 9AB}
  % \country{UK}
}
\email{mob@inf.ed.ac.uk}

\begin{abstract}

  Effective program synthesis requires a way to minimise the number of candidate
  programs being searched. A type signature, for example, places some small
  restrictions on the structure of potential candidates. We introduce and
  motivate a distilled program synthesis problem where a type signature is the
  only machine-readable information available, but does not sufficiently
  minimise the search space. To address this, we develop a system of property
  relations that can be used to flexibly encode and query information that was
  not previously available to the synthesiser. Our experience using these tools
  has been positive: by encoding simple properties and by using a minimal set of
  synthesis primitives, we have been able to synthesise complex programs in
  novel contexts.

\end{abstract}

\maketitle

\section{Motivation}

Program synthesis addresses the problem of \emph{automatically} generating
correct programs, rather than writing them by hand. A specification constrains
the allowable behaviour of a solution; the structure of generated programs and
their specifications varies between different synthesis problems.

Our central research aim is to better exploit heterogeneous hardware and
libraries in user code. Ideally, the compiler should automatically match user
code to the best compatible device or library. Our methodology is to
automatically learn a formal model of library behaviour that can be used to
discover compatible code. We do this by applying program synthesis to functions
in libraries with a C interface, which gives us the specification:
\begin{quote}
  Given a type signature $ {T \triangleq (\tau_0, \dots, \tau_n) \to
  \tau_r} $ and a function $ f : T $, a correct candidate program $ g : T $ is
  one that for all correctly typed lists of input arguments $ \mathbf{x} $, $
  f(\mathbf{x}) = g(\mathbf{x}) $.
\end{quote}

In the context of this problem, we are limited to the C type system for
compatibility with these existing library interfaces: we have a set of concrete
types (\mintinline{C}{int}, \mintinline{C}{float}, etc.), along with type
constructors for pointers (\mintinline{C}{int*}), aggregates
(\mintinline{C}{struct{int x; int y;}}) and arrays (\mintinline{C}{int[10]}).

Unfortunately, the synthesis problem corresponding to this specification is
intractable given only the type signature and behaviour of a function. The space
of potential solutions is too large to perform any kind of practical
search---we need some way of reducing its size. Oracle-guided inductive
synthesis \cite{Jha2010,Jha2015} makes use of additional information provided by
the function oracle to do this (for example, by providing minimal
counterexamples to a possible solution), but our black-box C functions do not
lend themselves well to this model.

The only formal description we have available for a library function is its type
signature, but informal (i.e.\ human- rather than machine-readable) information
can also be used to help the synthesiser. Special-casing individual sources of
information in the synthesiser is not scalable---ideally, it would be encoded
more formally alongside the type signature to allow the synthesiser to make
general use of them during the synthesis process.

% How, then, can this informal human-readable information be encoded such that a
% program synthesiser can take advantage of it? If we have such a representation,
% how usable and effective is it, and what programs can be synthesised that
% previously could not be?

% This leads us to the questions addressed by the remainder of this paper: how can
% informal, human-readable information about a function's behaviour and invariants
% be combined with its type signature to improve the synthesis process? Then,
% given such a representation, how usable and effective is it, and what programs
% can be synthesised that previously could not be?

\section{Property Relations}

The solution we propose is to augment function type signatures with a set of
property relations over the function's parameters and a set of literal values.
This representation is in the vein of a simple logic programming language and
allows for flexible encodings of human knowledge.

Our formal definition of these relations is as follows: Let $ f $ be a function
with type signature $ T \triangleq (\tau_0, \dots, \tau_n) \to \tau_r $, taking
parameters $ (p_0 : \tau_0, \dots, p_n : \tau_n) $.  Then, define:
\begin{align*}
  P & \triangleq \{p_0, \dots, p_n \} \\
  C & \triangleq \text{set of all C types } \\
  S & \triangleq \text{set of all string literals} \\
  N & \triangleq \text{set of all numeric literals} \\
  U & \triangleq P \cup C \cup S \cup N
\end{align*}
$ f $ is then associated with a set of relations $ R_f $. Each relation $ r_i
\in R_f $ satisfies $ r_i \subseteq U^k $ for some $ k > 0 $, and has a uniquely
identifying name $ I(r_i) \in S $ associated with it.

Less formally, named relations group sets of ``atoms'', where those atoms can be
function parameters, literal values or C types. Relations are associated with a
function and its type signature, and no particular semantics is attached to
them initially (the synthesiser supplies an interpretation for the
relations it is given for a function).

Our specification for these relations is intentionally simple---it is the
smallest definition we found that would allow for sufficiently useful properties
to be encoded. Additionally, a close relationship to the function's type
signature is maintained: both the signature and the associated relations express
facts about a function's behaviour and the meaning attached to its parameters.

\section{Synthesis and Queries}

The core methodology our program synthesiser is built on is
\emph{component-based} synthesis \cite{Lustig2009,Gulwani2011a}, where candidate
programs are composed from libraries of smaller fragments. A full description of
our synthesis algorithm is outside the scope of this paper, but a brief summary
(without considering optimisations or search strategies) is as follows:
\begin{itemize}
  \item First, a set of potential program fragments is assembled, using the
    type signature and property annotations to select ones most likely to be
    present in a correct program.
  \item Then, an iterative-deepening search enumerates valid compositions of
    fragments. Some fragments do not compose with others, depending on their
    context.
  \item For each composition, instruction sequences are sampled at set program
    locations specified by the fragment. Each resulting program is JIT-compiled
    and tested against the reference function.
\end{itemize}

This method is in the spirit of \emph{two-phase} sketching synthesis
\cite{Wang2017}, where an abstract or partial solution is synthesised first, and
is then instantiated to create a full solution.

Given a set of relations associated with a type signature, the synthesiser uses
a library of general heuristic patterns to bias its search towards more likely
programs. The synthesiser contains a set of fragment ``templates'', which are
partial programs parameterised on values $ \in U $ (as defined above), along
with rule-based heuristics for their instantiation, written using a simple query
language. Fragments provide a control flow structure that can have further
fragments nested inside it, or some specific sequence of data-flow instructions
likely to occur in a solution.

The queries used to govern fragment instantiation follow the style of a simple
logic programming language: a matching expression $ r(X, Y) $ is satisfied if
the relation named $ r $ in the current set is present, and contains a pair of
values that can be unified to the variables $ X, Y $. For a single match this is
trivial; but conjunctions may be formed, leading to more complex unifications.
Additionally, negative matches can be used (but require a conjunction with a
positive match to unify). Finally, a standard set of queries can be made of the
function's type signature. An example rule for instantiating a joint iteration
is:
\begin{align*}
  & size(X, N) \; \land size(Y, N) \; \land \\
  & type(X, T) \; \land type(Y, S) \; \land \; type(N, \mathtt{int}) \\
  & is{\text -}pointer(T) \land is{\text -}pointer(S) \\
  \Longrightarrow & \; \mathtt{zip\_loop}(N, T, X, S, Y)
\end{align*}

\section{Worked Example}

An illustrative example of how our approach helps to model and understand
performance libraries is the BLAS \cite{Blackford2001} standard. By synthesising
equivalent programs to functions in BLAS, we have been able to discover
opportunities for better library usage in existing scientific code.

The GEMV function in BLAS performs a general matrix-vector multiplication ($
\mathbf{y} \gets \alpha \mathbf{A} \mathbf{x} + \beta \mathbf{y} $). It is a
challenging target for program synthesisers: it contains nested control flow, 7
input parameters and complex array indexing expressions. For the signature:
\begin{minted}{c}
void gemv(int m, int n, float alpha, float *a,
          float *x, float beta, float *y);
\end{minted}
we provide the minimal set of property annotations:
\begin{align*}
  \{ size(x, n), size(y, m), output(y) \}
\end{align*}

Our synthesiser matches a number of general rules against these properties to
reach a correct solution for GEMV. The first is a general loop rule, which
matches both \mintinline{C}{x} and \mintinline{C}{y}:
\begin{align*}
  & size(X, N) \land type(N, int) \land type(X, T) \\ 
  & \land is{\text -}pointer(T) \Longrightarrow \mathtt{loop}(N, T, X)
\end{align*}

Two loops are instantiated and added to the set of potential fragments. The next
rule is one to perform a store to elements of \mintinline{C}{y}:
\begin{align*}
  output(X) \land type(X, T) \Longrightarrow \mathtt{store}(X, T)
\end{align*}

Finally, a negative-match rule matches \mintinline{C}{a}:
\begin{align*}
  & \overline{size(X, \_)} \land type(X, T) \land is{\text -}pointer(T) \\
  & \Longrightarrow \mathtt{affine\_access}(X, T)
\end{align*}

Other rules are present in the synthesiser, but these are the only ones that
match the properties associated with GEMV. The fragments instantiated by the
successful matches can compose in such a way that the algorithmic core of GEMV
is realised. After this, all that remains is to search for the correct sequence
of dataflow instructions:

\vspace{0.2cm}
\begin{center}
  \begin{minipage}{0.6\columnwidth}
    \begin{minted}{C}
for(int i = 0; i < m; ++i) {
  // code...
  for(int j = 0; j < n; ++j) {
    float v = x[?];
    // code...
  }
  y[i] = ?;
}
    \end{minted}
  \end{minipage}
\end{center}
\vspace{0.2cm}

Our synthesis results are promising: by using minimal property annotations
obtained from documentation, together with general-purpose heuristic queries, we
have been able to synthesise a wide variety of programs. The synthesised
programs come from a number of domains, and have led (with other work) to
significant performance improvements on real world scientific programs.

\vfill

\pagebreak

\bibliography{references.bib}

\end{document}